\documentstyle[aps]{revtex}

\newcommand{\bra}[1]{\left\langle #1 \right|}
\newcommand{\ket}[1]{\left| #1 \right\rangle}

\begin{document}

\title{Convergence properties of the effective interaction}

\author{P.J.\ Ellis}
\address{School of Physics and Astronomy, University of Minnesota,
Minneapolis, Minnesota 55455, USA}

\author{T. Engeland},
\address{Department of Physics,
University of Oslo, N-0316 Oslo, Norway}

\author{M. Hjorth-Jensen}
\address{Department of Physics,
University of Oslo, N-0316 Oslo, Norway}

\author{A. Holt}
\address{Department of Physics,
University of Oslo, N-0316 Oslo, Norway}

\and

\author{E. Osnes}
\address{Department of Physics,
University of Oslo, N-0316 Oslo, Norway}

\maketitle

\begin{abstract}
The convergence properties of
two perturbative schemes to sum the so-called folded diagrams are
critically reviewed in this work, with an emphasis on the
intruder state problem. The methods we study are the
approaches of Kuo and co-workers and Lee and Suzuki.
The suitability of the two schemes for shell-model
calculations is discussed.
\end{abstract}

\section{Introduction}

Recent works on the theory of the effective interaction
for finite nuclei, have focussed on
properties like hermiticity and the order-by-order
convergence of the perturbative
expansion for the effective interaction [1-4].
Especially, techniques to sum up subsets of diagrams to infinite
order have received increased interest. Two such schemes are
the summation of folded diagrams by
Kuo and co-workers, exposed
in ref.\ \cite{ko90} (hereafter referred to as the FD method),
and the approach proposed by Lee and Suzuki \cite{ls80}
(hereafter referred to as the LS method).
The starting point for these iterative schemes is
to define the so-called $\hat{Q}$-box, which in a diagrammatic
language means the sum of all valence-linked and irreducible
diagrams to a given order in the interaction. The $\hat{Q}$-box depends
on the so-called starting energy, defined as the unperturbed energy
of the interacting nucleons.

The purpose of these methods is to calculate an
effective interaction $H_{\mathrm{eff}}$
in order to reduce the full shell model eigenvalue problem,
to one which
is tractable within a physically selected space, the so-called
model space. One then obtains a secular equation
\begin{equation}
PH_{\mathrm{eff}}P\ket{\Psi}= E_P P\ket{\Psi},
\end{equation}
acting solely within the model space, defined by the projection operator
$P$, which projects out the model-space
components of the true eigenfunction
$\ket{\Psi}$. The quantity $E_P$ is the model-space
eigenvalue, and the $d$ eigenvalues $E_P$ ($d$ being the
dimension of the model space) are supposed to
reproduce the corresponding
exact eigenvalues \cite{ko90}.
The degrees of freedom not represented by the model
space are supposed to be taken care of by the perturbative expansion.
The excluded degrees of freedom are obtained through use
of the projection operator $Q$, with $P+Q=1$ and $PQ=0$.
The corresponding eigenvalues are $E_Q$.

The above techniques have been applied extensively to light nuclei,
in particular to the nuclei $^{18}$O and  $^{42}$Ca
where the  model space is rather simple,
two particles in the $sd$- or $pf$-shells, respectively.
Here, however, a special problem occurs, in the literature  called
the intruder-state problem. In short, this means that the $P$-
and $Q$-states do not separate. Experimentally, it is rather
well established
that the first excited 0$^{+}$ state in both $^{18}$O and  $^{42}$Ca is
dominated by $Q$-state components,
i.e. shell model configurations  outside
the $sd$- or $pf$-shell.
Then a comparison between theory and experiment becomes difficult.
Using the FD method to calculate $H_{\mathrm{eff}}$,
eq.\ (1) should reproduce the experimental levels  dominated by
the $P$-state configurations. For the two examples mentioned above
this means that the first excited $0^{+}$ state in  both
$^{18}$O and  $^{42}$Ca
should not be accounted for by the theory. On the other hand
the LS method,
for a properly chosen starting energy,
 should be able
to reproduce the  $d$ lowest  states of e.g., $^{18}$O and  $^{42}$Ca,
irrespectively of whether or not they are dominated by
$Q$-space components. In view of this, one might
expect the FD effective
interaction to be more appropriate
for shell-model calculations than the LS one.
An important application of such an effective interaction is
spectroscopic shell-model analyses of systems with more than two valence
particles, and then only $P$-space degrees of freedom are considered.

In several papers, e.g., ref.\ \cite{heho92},
we have investigated the above
mentioned nuclei using both the FD and the LS methods.
Our results show little difference in the final energy spectra.
The question then arises how one should relate the energy levels
calculated by respectively the FD and LS methods to the experimental
levels.

In the present paper we investigate this problem in more detail
and  explore the convergence properties of the FD and LS methods
in the presence of so-called intruder states.
In the next section  we employ a simple $2\times 2$ model
to study the difference between calculations with a complete and
a limited
perturbation expansion.
Then in section 3
we apply the FD and LS methods to  the nucleus $^{18}$O,
in order to test if differences seen in the exactly solvable
model pertain to the realistic case as well. Of importance
here is the fact that the simple $2\times 2$ model allows
us to define an exact $\hat{Q}$-box, whereas in the realistic
cases the $\hat{Q}$-box must be approximated to a given order
in the interaction. Finally, our conclusions are drawn in
section 4.

\section{Intruder states in an exactly solvable model}

It is a well-known fact that the presence of so-called
intruder states \cite{ko90,sw72}
may lead to divergence of the order-by-order
pertubative expansion
for $H_{\mathrm{eff}}$.

One way to handle intruder states is that of introducing an
enlarged
model space which includes intruder configurations.  Such calculations
do however become prohibitively time-consuming for nuclei heavier
than $sd$-shell nuclei.
There are also other approaches which aim at overcoming the
divergence of the order-by-order
expansion of $H_{\mathrm{eff}}$. One may,
e.g.,
regroup the perturbative expansion and perform infinite partial
summations to
obtain convergence. Two such possibilities are
represented by
the summation of the folded
diagrams by the FD and LS schemes discussed
above.
To better understand the above convergence
arguments and the structure of the folded-diagram and
the Lee-Suzuki methods we will
first demonstrate certain properties of these methods
by considering a simple $2\times 2$
model, employed e.g.\ by the authors of ref.\ \cite{eo77}
in their discussion of
perturbative many-body approaches.
Moreover, the model can be used to
demonstrate the connection between intruder states and the
convergence of the perturbative expansion. Last but not least, this
model allows us to define an exact $\hat{Q}$-box, which in turn can be
used to make statements about the convergence of the FD and LS
iterative schemes.

However, first
we need to repeat  the equations pertinent to the FD and LS methods.
The effective interaction for the FD method is
given by
\begin{equation}
      H_{\mathrm{eff}}=H_{0}+\lim_{n \rightarrow \infty}
     V_{\mathrm{eff}}^{(n)},
\end{equation}
with
\begin{equation}
      V_{\mathrm{eff}}^{(n)}={\displaystyle\sum_{m=0}^{\infty}}
      \hat{Q}_m\left\{
      V_{\mathrm{eff}}^{(n-1)}\right\}^{m}. \label{eq:fd}
\end{equation}
Here we have defined
$\hat{Q}_{m}=\frac{1}{m!}\frac{d^{m}\hat{Q}}
{d\omega^{m}}$.
The energy $\omega$ is the so-called starting energy,
defined as the unperturbed energy of the interacting
nucleons.

The Lee and Suzuki (LS)
expansion for the effective interaction is given as
\cite{ls80}
\begin{equation}
      H_{\mathrm{eff}}=H_{0}+\lim_{n \rightarrow \infty}R_{n},
      \label{eq:ls}
\end{equation}
with
\begin{equation}
     R_{n}=\left[1-\hat{Q}_{1}-\sum_{m=2}^{n-1}\hat{Q}_{m}
     \prod_{k=n-m+1}^{n-1}R_{k}\right]^{-1}\hat{Q}.
\end{equation}

Eqs.\ (\ref{eq:fd}) and (\ref{eq:ls}) differ not only in structure, but
also in convergence properties.
The convergence
criterion for the method of Lee and Suzuki is related to the
choice of the so-called starting energy
$\omega$  {\em only}.
The LS expansion converges to those eigenvalues which are
closest to $\omega$, which means that we should be able even to
reproduce the $Q$-space eigenvalues given
an appropriate starting energy.
This means that if the LS expansion converges for the $0_2^+$
state of, e.g.\ $^{18}$O,
the structure of the exact wave function
does not correspond to that of the two-particle space we have chosen.
In contrast,
the FD-method converges to the states dominated
by the model-space components.

In order to study the importance of intruder states, we let
our hamiltonian depend linearly on a strength parameter $z$
\[
       H=H_0+zH_1,
\]
with $0\leq z\leq1$, where the limits $z=0$ and $z=1$ represent the
non-interacting (unperturbed) and fully interacting system, respectively.
The model is an eigenvalue
problem with only two available states, which we label
$P$ and $Q$. Below we will let
state $P$ represent the model-space
eigenvalue whereas state $Q$ represents
the eigenvalue of the excluded space.
The unperturbed solutions to this problem are
\begin{equation}
       H_0\Phi_P =\epsilon_P\Phi_P
\end{equation}
and
\begin{equation}
       H_0\Phi_Q =\epsilon_Q\Phi_Q,
\end{equation}
with $\epsilon_P < \epsilon_Q$. We label the off-diagonal
matrix elements $X$, while $X_P=\bra{\Phi_P}H_1\ket{\Phi_P}$ and
$X_Q=\bra{\Phi_Q}H_1\ket{\Phi_Q}$.
The exact eigenvalues problem
\begin{equation}
\left(\begin{array}{cc}\epsilon_P+zX_P &zX \\
zX &\epsilon_Q+zX_Q \end{array}\right)
\end{equation}
yields
\begin{eqnarray}
     \label{eq:exact}
     E(z)=&\frac{1}{2}\left\{\epsilon_P +\epsilon_Q +zX_P
     +zX_Q \pm \left(
     \epsilon_Q -\epsilon_P +zX_Q-zX_P\right) \right. \\ \nonumber
     & \left. \times\sqrt{1+\frac{4z^2X^2}{\left(
     \epsilon_Q -\epsilon_P +zX_Q-zX_P\right)^2}}
     \right\}.
\end{eqnarray}
The authors of ref.\ \cite{eo77}
demonstrated how Brillouin-Wigner and
Rayleigh-Schr\"{o}dinger (RS) perturbation theories relate
within the framework of this simple model.
An RS expansion for the lowest
eigenstate (defining states $P$ and $Q$ as the model and excluded
spaces, respectively) can be obtained by expanding the lowest
eigenvalue as
\begin{equation}
      E=\epsilon_P +zX_P+\frac{z^2X^2}{\epsilon_P -\epsilon_Q}+
      \frac{z^3X^2(X_Q-X_P)}{(\epsilon_P -\epsilon_Q)^2}+
      \frac{z^4X^2(X_Q-X_P)^2}{(\epsilon_P -\epsilon_Q)^3}
      -\frac{z^4X^4}{(\epsilon_P -\epsilon_Q)^3}+\dots,
      \label{eq:modela}
\end{equation}
which can be viewed as an effective interaction for state $P$ in which
state $Q$ is taken into account to successive orders of the perturbation.
In this work we choose the parameters $\epsilon_P=0$, $\epsilon_Q=4$,
$X_P=-X_Q=3$ and $X=0.2$. The exact solutions
given by eq.\ (\ref{eq:exact})
are shown in fig.\ 1 as functions of the
strength parameter $z$. Pertinent to our choice of
parameters, is that at $z\geq 2/3$,  the lowest eigenstate is
dominated by $\Phi_Q$ while the upper is $\Phi_P$. At $z=1$ the
$\Phi_P$ mixing of the lowest eigenvalue
is $1\%$ while for $z\leq 2/3$
we have a $\Phi_P$ component of more than $90\%$.
The character of the eigenvectors has therefore been interchanged
when passing $z=2/3$. The value of the parameter $X$ represents the
strength of the coupling between the model space and the excluded space.
Thus, this simple
model allows us to study how the perturbation expansion with a
model space defined to consist of state $P$ only, behaves
as the interaction strength $z$ increases. The order-by-order
convergence in eq.\ (\ref{eq:modela}) was discussed by the authors
of ref.\ \cite{eo77}. Here we will thence only repeat
their conclusion:
For small values of $z$ one obtains
good convergence to the lower eigenvalue. For larger values of $z$
(i.e.\ $z> 2/3$),
increasing orders in the perturbation expansion yield a divergent
perturbation series, as expected \cite{sw72}.
However, it may be possible to rewrite
the perturbative expansion in such a way that one sums
subsets of diagrams to all orders. The hope is then that the
expansion becomes convergent for appropriate infinite partial
summations.
This is actually the philosophy
behind both the FD method and the LS method. Having defined
a set of linked and irreducible valence diagrams,
the so-called $\hat{Q}$-box\footnote{
The $\hat{Q}$-box should not be confused with the exclusion
operator $Q$.}, we can define an
iterative scheme to sum the contributions
from folded diagrams.
The $\hat{Q}$-box serves therefore as the starting point
for our iterative schemes, and within the framework of our
$2\times 2$ model we can
study the FD and LS methods for various
approximations of the $\hat{Q}$-box.
As an example, to fifth order in the parameter $z$ we have
\begin{equation}
       \hat{Q}_5=zX_P+\frac{z^2X^2}{\epsilon_P -\epsilon_Q}+
       \frac{z^3X^2X_Q}{(\epsilon_P -\epsilon_Q)^2}+
       \frac{z^4X^2X_Q^2}{(\epsilon_P -\epsilon_Q)^3}+
        \frac{z^5X^2X_Q^3}{(\epsilon_P -\epsilon_Q)^4},
       \label{eq:qapprox}
\end{equation}
and it
is easy to see that a $\hat{Q}$-box of order $l+2$ can be written
as
\[
               \hat{Q}_{l+2}=
                zX_P+\frac{z^2X^2}{\epsilon_P -\epsilon_Q-zX_Q}
                \left\{1-\left(\frac{zX_Q}
                {\epsilon_P-\epsilon_Q}\right)^{l+1}\right\},
\]
which in the limit $l\rightarrow \infty$
gives\footnote{$l=0$ gives a second-order $\hat{Q}$-box, $l=1$ a third-order
$\hat{Q}$-box and so forth.}
\begin{equation}
      \hat{Q}_{\mathrm{exact}}(\epsilon_P)=
      zX_P+\frac{z^2X^2}{\epsilon_P -\epsilon_Q-zX_Q},
      \label{eq:qexact}
\end{equation}
if
\begin{equation}
     \left|\frac{zX_Q}{\epsilon_P-\epsilon_Q}\right| < 1.
     \label{eq:constr}
\end{equation}
The latter equation clearly restricts the possible
values of $\epsilon_P$ for
given $X_Q$ and $\epsilon_Q$. Actually,
if we let $\epsilon_P$ vary, our choices
for $X_Q$ and $\epsilon_Q$ restrict $\epsilon_P$ to
$\epsilon_P \leq 1$ and $\epsilon_P\geq 7$, in order
to have a finite $\hat{Q}$-box.

The exact $\hat{Q}$-box
may be used to define the first iteration of
the FD expansion as a function of the starting energy
$\omega$ as\footnote{We replace here $\epsilon_P$ with $\omega$.}
\begin{equation}
      \lambda_1= \omega+\hat{Q}_{\mathrm{exact}}(\omega)
\end{equation}
The subsequent steps are
\begin{equation}
     \lambda_2= \lambda_1+
     {\displaystyle \sum_{m=1}^{\infty}
     \frac{1}{m!}\frac{d^m \hat{Q}}{d\omega^m}(\lambda_1
     -\omega )^m },
\end{equation}
and
\begin{equation}
    \lambda_3= \lambda_1+
     {\displaystyle \sum_{m=1}^{\infty}
     \frac{1}{m!}\frac{d^m \hat{Q}}{d\omega^m}(\lambda_2-
     \omega )^m }.
\end{equation}
In general we have
\begin{equation}
     \lambda_n= \lambda_1+
     {\displaystyle \sum_{m=1}^{\infty}
     \frac{1}{m!}\frac{d^m \hat{Q}}{d\omega^m}(\lambda_{n-1}-
     \omega )^m }.
\end{equation}
If this iteration scheme converges we have, with
$\lambda=\lambda_{n}=\lambda_{n-1}$,
\begin{equation}
   \lambda=\omega +zX_P+\frac{z^2X^2}{\lambda -\epsilon_Q-zX_Q},
   \label{eq:exact2}
\end{equation}
which is just eq.\ (\ref{eq:exact}), so that we have
obtained the true eigenvalues.
In a similar way we can use the LS expansion defined
in eq.\ (\ref{eq:ls}) to sum the folded diagrams.
\begin{table}[hbtp]
\begin{center}
\caption{The exact solutions $E_P$ ( model space) and
$E_Q$ ( excluded space) of eq.\ (9)
as functions of the strength parameter $z$.
The results obtained with the LS and FD methods with an
exact $\hat{Q}$-box as functions
of the starting energy $\omega$ are also given.}
\begin{tabular}{rrrrrr}
&&&&&\\\hline
\multicolumn{1}{c}{$\omega$}&
\multicolumn{1}{c}{$z$}&
\multicolumn{1}{c}{$E_P$}&
\multicolumn{1}{c}{$E_Q$}&
\multicolumn{1}{c}{$FD$}&
\multicolumn{1}{c}{$LS$}\\
\hline
0.5&0.0&0.00&4.00&0.00&0.00\\
   &0.2&0.60&3.40&0.60&0.60\\
   &0.4&1.20&2.80&1.20&1.20\\
   &0.6&1.77&2.23&1.77&1.77\\
   &0.7&2.17&1.83&2.17&1.83\\
   &0.8&2.43&1.57&2.43&1.57\\
   &1.0&3.02&0.98&3.02&0.98\\
7.5&0.0&0.00&4.00&0.00&4.00\\
   &0.2&0.60&3.40&0.60&3.40\\
   &0.4&1.20&2.80&1.20&2.80\\
   &0.6&1.77&2.23&1.77&2.23\\
   &0.7&2.17&1.83&2.17&2.17\\
   &0.8&2.43&1.57&2.43&2.43\\
   &1.0&3.02&0.98&3.02&3.02\\
   \hline
\end{tabular}
\end{center} \label{tab:table1}
\end{table}
We demonstrate the properties of the LS and FD methods with an
exact $\hat{Q}$-box in table 1 for two starting energies,
$0.5$ and $7.5$ (arbitrary units). Clearly, we see that with a starting
energy $0.5$ the LS method yields the $Q$-space eigenvalue at $z\geq 2/3$.
Below $z=2/3$, the LS method reproduces the $P$-space eigenvalue.
With a starting energy of $7.5$, the LS method
gives the $Q$-space eigenvalue
for $z\leq 2/3$, whereas the $P$-space eigenvalue is reproduced for
$z\geq 2/3$. Thus, this simple model demonstrates nicely the properties of
the LS scheme, i.e.\ it converges to those eigenvalues which are
closest to the chosen starting energy, irrespectively of the structure of
the wave function. From table 1 we also see that the FD
method always reproduces the $P$-space eigenvalue.
At $z=2/3$, the FD
scheme does not stabilize, the eigenvalue fluctuates and there is
no convergence. More precisely this means that we can not
go from the lower to the upper
eigenvalue along the real axis $z$ \cite{eo77}.
The fluctuation is intimately connected to the convergence
criterion for the FD scheme. If one of the solutions contains
more than $50\%$ valence state intensity, the FD method converges
to that solution. At $z=2/3$, equal admixtures of $\Phi_P$
and $\Phi_Q$ occur in the true wave functions.

This simple example, {\em starting with the
exact $\hat{Q}$-box},  serves to demonstrate significant
differences in convergence behavior of the two methods. Of
importance is the fact that with the LS scheme
we are able to reproduce a
$Q$-space state insofar we define a starting energy which is
close to the actual $Q$-space state.
However, in actual nuclear structure calculations, we are not
able to define an exact $\hat{Q}$-box. The question we wish
to bring to the attention in this work, is if an
approximation of the $\hat{Q}$-box still gives the same
difference between the LS and FD methods at $z=1$.
To shed light on this, we exhibit in
fig.\ 1 results obtained with various low-order
approximations to the $\hat{Q}$-box, recall eq.\ (\ref{eq:qapprox}).
The starting energy for the LS and
FD calculations was set equal $0.5$, since
we are interested in seeing if the convergence criterion for the
LS method holds for an approximate $\hat{Q}$-box.
Clearly, with a $\hat{Q}$-box of second order
in the interaction, the
difference\footnote{It ought be emphasized that all results
with either the FD or the LS method represent converged
eigenvalues, which means that contributions from folded diagrams
to high order are included.}
between the LS and the FD method is negligible.
A second-order $\hat{Q}$-box corresponds to setting $X_Q=0$, which
means that in the range of {\em physically interesting}
$z$ values ($0\leq z \leq 1$)
there is no intruder state, yielding an approximately straight line
for the LS method.
Up to fourth order in the interaction, both
the LS and FD methods yield almost
the same value. With a fifth-order
$\hat{Q}$-box the FD method becomes unstable at $z=1$.
This can
be inferred from the structure of the FD expansion,
which with a fifth-order $\hat{Q}$-box
reads
\[
     \lambda_n=\epsilon_P+zX_P+\frac{z^2X^2}{\lambda_{n-1} -\epsilon_Q}+
      \frac{z^3X^2X_Q}{(\lambda_{n-1} -\epsilon_Q)^2}+
      \frac{z^4X^2X_Q^2}{(\lambda_{n-1} -\epsilon_Q)^3}+
     \frac{z^5X^2X_Q^3}{(\lambda_{n-1} -\epsilon_Q)^4}.
\]
The first iteration is just $\lambda_1=\epsilon_P+\hat{Q}$,
which gives a result close to the model space eigenvalue
$\approx 3$ at $z=1$. Inserting $\lambda_1$ and higher iterations
in the above expansion
results in an
FD expansion which for each iteration will fluctuate
between a large value $\lambda$ and $\epsilon_P+X_P$.
The large eigenvalue stems
from the fact that the energy denominators in the higher-order terms become
small. With $X=0.2$ this oscillating behavior sets inn already at fifth
order in the interaction, whereas
if we choose the coupling
between the model space and the excluded space to be $X=0.01$,
this divergence appears first
with a $\hat{Q}$-box of tenth order. However,
we will ultimately end up with a series which fluctuates,
except for the trivial case $X=0$.
With the given parameters and a $\hat{Q}$-box of tenth or higher
order in the interaction,
the FD method converges only if $z<2/3$.
If the FD method converges, we obtain a
result close to the model-space eigenvalue,
and, if the coupling between the model space and
the excluded space is weak,
low-order perturbation theory works rather well.
\begin{figure}[hbtp]
      \setlength{\unitlength}{1mm}
      \begin{picture}(140,150)
      \end{picture}
       \caption{
        Exact solutions (solid line) and the results
        obtained with  the FD and LS methods
        with various approximations to the
        $\hat{Q}$-box. FD-2nd and LS-2nd (dashed line)
        indicate that the FD and LS methods
        were used with a second-order $\hat{Q}$-box,
        while FD-4th and LS-4th are the results obtained
        with a fourth-order $\hat{Q}$-box. Similarly, LS-10th and LS-50th
        are the LS results with a tenth- and fiftieth-order
        $\hat{Q}$-box, respectively.
        A starting energy $0.5$ was chosen in all calculations.}
        \label{fig:model}
\end{figure}
For the LS method we note that with a low-order $\hat{Q}$-box we are
not able to reproduce the lowest eigenstate, even with an adequately
chosen starting energy, as stated by the convergence criterion
of the LS method.  We actually have to go as far as to
a $\hat{Q}$-box of fiftieth order before we get a result close
to the lowest eigenvalue. However, if the order of the $\hat{Q}$-box
is further increased ($\approx 150$ with the above parameters), even
the LS method diverges.
Thus, in summary, the general convergence properties of
the LS method discussed in the
introduction are demonstrated only for an exact $\hat{Q}$-box.
If the $\hat{Q}$-box is approximated to a given order in the
interaction, the convergence criterion cannot be applied.

The implications of these results for the more realistic
cases will be discussed in the next section.

\section{Realistic effective interactions}

In calculations of effective interactions for $^{18}$O,
one normally chooses a model space
consisting of a closed $^{16}$O core with two nucleons
in the single-particle states of the $sd$-shell.
This model space consists of two-particle states only,
and for $J=0$ we have
three $0^+$ states.
However, from model calculations of energies and
 electromagnetic properties and experimental
data, the belief is that the $0_2^+$ state is predominantly of a
four-particle-two-hole nature, while the ground
state and the $0_3^+$ state are classified as two-particle
states \cite{ee70}.

Before we proceed to study the convergence behavior of
the folded-diagram and Lee-Suzuki
expansions, let us first show how one can
obtain useful information about the structure of the
eigenstates from the slopes
of the eigenvalues as functions of the starting energy $\omega$.
Using the property $P+Q=1$, we can rewrite the energy dependent
eigenvalue problem as
\begin{equation}
     \left(\begin{array}{cc}PHP &PHQ\\QHP& QHQ\end{array}\right)
     \left(\begin{array}{c} P\Psi\\Q\Psi\end{array}\right)
     =E \left(\begin{array}{c}P\Psi\\Q\Psi\end{array}\right),
     \label{eq:fulle}
\end{equation}
where $E$ is the exact eigenvalue.
{}From eq.\ (\ref{eq:fulle}) we obtain
\begin{equation}
     E\left\langle P\Psi |  P\Psi\right\rangle =
     \bra{P\Psi}PHP+PHQ\frac{1}{\omega - QHQ}QHP\ket{P\Psi},
    \label{eq:graphsol}
\end{equation}
with $\omega=E$ and
\begin{equation}
     \left\langle Q\Psi | Q\Psi\right\rangle =
     \bra{P\Psi}HQ\frac{1}{\left(\omega - QHQ\right)^2}QH\ket{P\Psi},
\end{equation}
such that
\begin{equation}
    \left\langle P\Psi | P\Psi\right\rangle =
    \frac{1}{1-\frac{dE}{d\omega}|_{\omega=E}}.
\label{eq:overlap}
\end{equation}
This equation states that the slopes at the
intersection between the r.h.s.\ of eq.\ (\ref{eq:graphsol})
and the line $E=\omega$
provide information about the wave functions.
The slopes $\frac{dE}{d\omega}$
should always satisfy
\[
   \frac{dE}{d\omega}\leq 0,
\]
in order to have $\left\langle P\Psi | P\Psi\right\rangle \leq 1$.
Using $H=H_0+H_1$ we can rewrite the r.h.s.\ of eq.\ (\ref{eq:graphsol})
as
\begin{equation}
    PHP\ket{\Psi}=
    PH_0P\ket{\Psi} + P\hat{Q}P\ket{\Psi}=EP\ket{\Psi},
\end{equation}
where we have introduced the $\hat{Q}$-box defined as
\begin{equation}
     P\hat{Q}P=PH_1P+
     PH_1Q\frac{1}{\omega-QHQ}QH_1P.
    \label{eq:qbox}
\end{equation}
How to evaluate the $\hat{Q}$-box is however an open question. Here
we will expand the denominator of eq.\ (\ref{eq:qbox}) such that
we obtain a perturbative expansion for the $\hat{Q}$-box which reads
\begin{equation}
     P\hat{Q}P=PH_1P+
     P\left(H_1 \frac{Q}{\omega-H_{0}}H_1+H_1
     \frac{Q}{\omega-H_{0}}H_1 \frac{Q}{\omega-H_{0}}H_1 +\dots\right)P.
\end{equation}
The $\hat{Q}$-box is made up of non-folded diagrams which are irreducible
and valence linked. A diagram is said to be irreducible if between each pair
of vertices there is at least one hole state or a particle state outside
the model space. In a valence-linked diagram the interactions are linked
(via fermion lines) to at least one valence line. Note that a valence-linked
diagram can be either connected (consisting of a single piece) or
disconnected, which in the case of a two-body interaction
means that we have two one-body diagrams. In the FD and LS expansions,
where we include folded diagrams as well, the
disconnected diagrams are found to cancel out \cite{ko90}.

We evaluate the
$\hat{Q}$-box by
including all two-body diagrams through third order in the
reaction matrix $G$, as described in ref.\ \cite{hom92}. The one-body
diagrams plus the unperturbed term $H_0$ are approximated by the
experimental single-particle energies.
Further, since we do not evaluate
diagrams with Hartree-Fock insertions, we have no disconnected diagrams
through third order in the interaction in our definition of the
$\hat{Q}$-box.
The $G$-matrix replaces the interaction term $H_1$ through $H_1=G-U$,
$U$ being an adequately chosen auxiliary potential, and
is obtained through the
solution of the Bethe-Goldstone equation
\begin{equation}
     G(\omega )=V+V\frac{\tilde{Q}}{\omega - H_0}G(\omega ),
     \label{eq:gmat}
\end{equation}
where $V$ is the free nucleon-nucleon interaction. Note that
the Pauli operator $\tilde{Q}$ may differ
from the definition of $Q$ in the above
perturbative expansions. Eq.\ (\ref{eq:gmat}) is solved by using the
so-called double-partition
scheme described in ref.\ \cite{kkko76}. For $V$, we take the
parameters of the
Bonn B potential as defined in table A.2 of ref.\ \cite{mac89} with
the Pauli operator $\tilde{Q}$ defined so as to prevent scattering into
states with one particle in the $0s$ or $0p$ shells or two particles
in the $1s0d$ or $1p0f$ shells. A harmonic oscillator single-particle
basis
was employed with an oscillator parameter of $1.72$ fm.
\begin{figure}[hbtp]
      \setlength{\unitlength}{1mm}
      \begin{picture}(140,150)
    \end{picture}
    \caption{The eigenvalues of eq.\ (20) for the three
    $0^+$ states of $^{18}$O.}
    \label{fig:fig2}
\end{figure}
In fig.\ 2 we show the graphical solutions for the $0_1^+$,
$0_2^+$ and
$0_3^+$ states.
As can be seen from fig.\ 2, the two lowest eigenstates
exhibit negative derivates in the region
near the graphical solutions, $-12.2$
and $-8.3$ MeV for the $0_1^+$ and the $0_2^+$ states, respectively.
Correspondingly, the model space overlaps of eq.\ (\ref{eq:overlap})
in the region $-20$ to $-4$
MeV for the starting energy is of the order $\approx 0.93-0.95$ for the
$0_1^+$ state and $\approx 0.98-0.99$ for the $0_2^+$ state. We do not
attempt to calculate the corresponding numbers near the crossing for the
$0_3^+$ state, since this would require the eigenvalue to be calculated
at positive starting energies, which is not possible since the
$G$-matrix is
calculated at negative energies only.
We are therefore not able to make any
statement about the wave function overlaps
for this state. In our further
discussion we will therefore omit this state.

For the $0_1^+$ state, the derivative
becomes positive first at $-28$ MeV ($\frac{dE}{d\omega}=0.285$), and the
eigenvalue diverges as we approach the poles $\omega = QH_0Q$.
However, it is interesting to
see that within our approximation to the $\hat{Q}$-box,
the $P$-space overlap is close to one for the starting energies of
interest, i.e.\ near the graphical solutions. This means that
the mixing from $Q$ space states is rather weak within our perturbative
scheme. As discussed above, the experimental
$0_2^+$ state is expected to be of
four-particle-two-hole structure (4p2h).
Although we include the second-order 4p2h diagram in our definition
of the $\hat{Q}$-box,
this contribution is not sufficient to represent
the structure of the 4p2h correlations of the $0_2^+$ state.
The 4p2h diagram
is also the only diagram in our third-order $\hat{Q}$-box which
may represent intruder state configurations.
The omission of the latter diagram in our calculations
does not change the qualitative
aspect of the above wave function overlaps, indicating that we have a
rather small $Q$-space component in conventional calculations
of the effective interaction in the region of the graphical solutions.
We may then summarize this to say
that for a perturbative expansion up to
third order in the $G$-matrix, the $Q$-space
component is weak and one should
identify the second eigenvalue of the model calculation with
the $0_3^+$ experimental state.

{}From the analysis in the previous section and the above
results, we expect
that the differences between the FD and the LS method, i.e.\
when comparing the converged results, should be small in the energy domain
of interest. In particular, the $P$-space overlaps for the
$0_1^+$ and the  $0_2^+$ should be close to unity in both
schemes.
In fig.\ 3 we plot the eigenvalues as functions of the
starting energy for the FD and LS
methods\footnote{The one-body contributions
plus $H_0$ are also here approximated with the experimental
single-particle
energies}.
\begin{figure}[hbtp]
      \setlength{\unitlength}{1mm}
      \begin{picture}(140,150)
     \end{picture}
     \caption{The eigenvalues derived with the FD and LS methods
              for the $0^+$ states of $^{18}$O.}
     \label{fig:fig3}
\end{figure}
We observe that for starting energies within the range of the
experimental values, the FD and LS methods show only small differences.
The results here should therefore be interpreted in the same way as
those exhibited in fig.\ 1. In that
figure we showed that for low-order $\hat{Q}$-boxes,
the FD and LS methods
gave essentially the same results. This leads
us in turn to state that the
convergence criterion for the LS method
{\em applies only if we are able to
define an approximately exact
$\hat{Q}$-box}. In realistic calculations this is however not the
case, although both series converge rather rapidly, almost irrespectively
of the choice
of $\hat{Q}$-box. Thus, in calculations of the spectra for, e.g.\
$^{18}$O,
we cannot choose a starting energy close to the $0_2^+$ state, in order
to reproduce that state with the LS method, since the
$\hat{Q}$-box does not incorporate important
$Q$-space configurations. Rather, the LS and FD methods
give similar results, and {\em the second $0^+$
state obtained from the calculation
should be identified with the experimental $0_3^+$ state,
since this is believed
to be predominantly a  two-particle state}.

Finally, an important subtlety arises. Let us assume that we are
able to define an exact $\hat{Q}$-box for the two-body interaction.
With an appropriately chosen starting energy, we should then be able to
reproduce the intruder state in e.g.\ $^{18}$O employing the LS method.
This interaction is then conventionally used in shell-model
calculations to obtain observables of systems with more than
two valence nucleons. However, for such heavier systems, the degrees
of freedom of the intruder state in e.g.\ $^{18}$O may not
be present. Thus, this two-body
interaction obtained with the LS method should not be used
in calculations of spectra of other $sd$-shell nuclei. With the
LS  method, we should therefore, in principle, recalculate
the $n$-body interaction for the actual nucleus with $n$ valence
particles.
This should be contrasted to the effective interaction obtained
with the FD method. The latter reproduces only those states which have
the largest $P$-space overlaps. If we restrict the attention to the
degrees of freedom represented by the model space, we can use
this two-body interaction in heavier systems as well.

\section{Conclusions}
We have critically reviewed the convergence properties
of the folded-diagram
(FD) expansion of Kuo and co-workers \cite{ko90} and the iterative
scheme of Lee and Suzuki \cite{ls80}. The behavior of these perturbative
expansions was first studied within the framework of an exactly solvable
model. Next, we investigated the convergence properties by applying the
above two methods to a test nucleus, $^{18}$O. Our conclusions are:
\begin{itemize}
\item The exactly solvable model allows us to define an exact
$\hat{Q}$-box. With an exact $\hat{Q}$-box we were
able to demonstrate that,
if we choose a starting energy close to a given eigenvalue, either
$P$- or $Q$-space eigenvalues, the LS expansion converges to that
eigenvalue.
The FD method
converges always to that eigenvalue which has the largest $P$-space
overlap.
\item Still within the exactly solvable model, we have shown that if one
chooses a $\hat{Q}$-box of low order in the interaction, the FD
and LS methods give almost similar results, irrespectively of
the choice of starting energy.
At low orders in the interaction we are not able to reproduce the
$Q$-space state with the LS method, even with an appropriately
chosen starting energy. The
FD method (if it converges) always gives a
result close to the  eigenvalue with the largest $P$-space
overlap. With increasing order of the interaction, the $\hat{Q}$-box
incorporates more $Q$-space degrees of freedom and the LS and FD methods
give unstable results.
\item In realistic calculations, like those for $^{18}$O shown here,
we can only define an approximate $\hat{Q}$-box, to third
order in the interaction in the present case. It is then seen that the
FD and LS methods yield almost the same results, in agreement with the
findings obtained from the simple model. Since the FD method
converges to those states which have the largest $P$-space overlap,
the structure of the LS wave function should therefore also have a large
$P$-space overlap.
With an approximative $\hat{Q}$-box,
we can therefore not choose a starting energy close to an intruder state
in order to reproduce this state with the LS method.
Thus, of the two methods, the FD
method is the preferable one
in nuclear structure calculations since
it converges to those eigenstates which have the largest $P$-space
overlap, a desirable property since we in general can only employ a
rather small shell-model space.
\end{itemize}

\end{document}